\documentclass[prd,showpacs]{revtex4}
\usepackage[dvipdfm]{hyperref,graphicx}
\usepackage{bm}
\begin{document}
\title{Sensitivity Parameter and  Time Variations of Fundamental Constants}

\author{Su Yan}
\affiliation{Department of Physics and Astronomy,
Northwestern University, Evanston, IL 60208, USA}
\begin{abstract}
The sensitivity parameter is widely used for quantifying fine tuning. However, examples show it fails to give correct results under certain circumstances. We argue that the problems of the sensitivity parameter are almost identical to the consequences we have to solved if the time-varying fine structure constant is proved to be true. The high sensitivity of the energy scale parameter \(\Lambda\) to the dimensionless coupling constant plays an important role in these problems. It affects the reliability of the sensitivity parameter via mechanisms such as dynamical symmetry breaking, chiral symmetry breaking etc. The reliability of the sensitivity parameter can be improved if it is used properly.

\end{abstract}

\maketitle

\section{Introduction}

Recent astrophysical observations have shown the possibility of small time variation of the fine structure constant on the cosmological time scale:
\begin{equation}
\frac{\delta\alpha}{\alpha}=(-0.57\pm0.11)\times10^{-5} 
\end{equation}
for the redshift range \(0.2<z<3.7\)\cite{Webb,Murphy}. Some researchers suggested that such time variations of the coupling constants could yield large time variations of other physical quantities\cite{Marciano,Calmet,Langacker,Dent}, which are difficult to explain under current high energy physics framework. For example, X.Calmet et al\cite{Calmet} and Langacker et al \cite{Langacker} suggested that because of the following relation between the QCD scale \(\Lambda_{QCD}\) and the strong coupling:
\begin{equation}
\frac{\dot{\Lambda}_{QCD}}{\Lambda_{QCD}}=\ln\left(\frac{\mu}{\Lambda_{QCD}}\right)\frac{\dot{\alpha}_s}{\alpha_s}\simeq 38\frac{\dot{\alpha}_s}{\alpha_s}
\label{eq2}
\end{equation}
those physical quantities related to the QCD scale could also exhibit large time variations.
For example, the proton mass can be linked to the QCD scale by\cite{Langacker}:
\begin{equation}
\frac{\dot{m_p}}{m_p}\simeq\frac{\dot{\Lambda}_{QCD}}{\Lambda_{QCD}}
\end{equation}
which will lead to large time variation of the proton mass. Theoretically it is difficult to explain
why such large variations are allowed. These consequences may become theoretical problems if the fundamental couplings are really time varying.

Similar results for other physical quantities can also be found\cite{Dent,Calmet,Langacker,Dine}. Although the mechanisms involved are highly model dependent, the results are almost identical: those physical quantities all exhibit large time variations. Understanding why the nature allows such large time variations is difficult, which still lacks acceptable explanations. However, the previous studies\cite{Dent,Calmet,Langacker,Dine} give us a clear picture of the relations between the variations of the couplings and the variations of the mass quantities, which will be very helpful for us to understand the problems of the sensitivity parameter.

The sensitivity parameter\cite{BG} proposed by R. Barbieri, G.F.Giudice and J. Ellis et al is a quantitative measure widely adopted to quantify the severity of fine-tuning\cite{Wilson}. Its idea is quite simple. Think of a weak scale measurable parameter \(y\) which is affected by the fine-tuning problem, it shows strong dependency on a fundamental Lagrangian parameter \(x\) at the Planck scale. The sensitivity parameter \(c(x)\) then can be defined as:

\begin{equation}
\frac{\delta y}{y}=c(x)\frac{\delta x}{x}
\label{eqx3}
\end{equation} 

Along with this definition, usually a particular value \(c=10\) is chosen as the maximum allowed value for the parameter \(x\) to be categorized as ``natural''\cite{BG}, though the choice of \(c=10\) itself is quite arbitrary. The measure \(c(x)\) along with the cut-off value \(c=10\) constitute the sensitivity criterion.

Although generally the largeness of the sensitivity parameter is in good correspondence with the severity of fine-tuning, it does fail under certain circumstances\cite{GWA,AC2,AC3,CS}. For instance, G. Anderson et al\cite{GWA} pointed out that if the proton mass \(m_p\) is expressed as the function of the strong coupling constant \(g\) and the Plank mass \(M_P\) by dimensional transmutation:

\begin{equation}
m_p \approx M_Pe^{-(4\pi)^2/bg^2(M_P)}
\label{eqx4}
\end{equation}
which yields the sensitivity parameter:
\begin{equation}
c(g)=\frac{4\pi}{b}\frac{1}{\alpha_s(M_P)}\gtrsim 100
\label{eqx5}
\end{equation}
even though the proton mass is quite ``natural".

Later P. Ciafaloni et al\cite{CS} also pointed out that when the Z-boson mass \(M_Z\) is dynamically determined through gaugino condensation in a ``hidden'' sector:
\begin{equation}
M_Z\approx M_P e^{-l/g^2_H}
\label{eqx6}
\end{equation}
where \(g_H\) is the hidden sector gauge coupling constant renormalized at \(M_P\), \(l\) is a constant. Again the sensitivity \(c\) of \(M_Z\) will always be much greater than the fine-tuning cut-off \(c=10\), although it is well known that Z-boson mass \(M_Z\) is not always fine-tuned. 

Similar examples can be found in many places\cite{BR,BS,GRS,RS,Castano}. These examples all show that the sensitivity criterion is problematic. Although previously many authors have attempted to explain these problems, yet none of them can claim quantitative rigor\cite{Abe:2001npb}. No widely accepted explanation has ever been proposed to explain why sometimes the sensitivities are so large for the quantities that are apparently nothing to do with fine-tuning. The problem of the sensitivity parameter is still an open question. 

With a careful exam of these cases, it is not difficult to find  that the sensitivity parameter fails only when a dimensionful mass parameter is compared with a dimensionless coupling constant. This scenario is very much similar to the problems related with the time-varying fine structure constant. To investigate their possible connections,
simply rewrite Eq.~\ref{eq2}, we have:
\begin{equation}
\Lambda_{QCD}=\mu e^{2\pi/\beta_0\alpha_s(\mu)}
\end{equation}
which is almost identical to Eq.~\ref{eqx4} and Eq.~\ref{eqx6}. The mathematical similarities remind us that these two types of problems might be identical theoretically.  Because of the importance of the fine-tuning problems in particle physics, it is worth to investigate the relationship between the sensitivity related problems and the time-varying coupling constant related problems. Then we can further understand the reason why the sensitivity parameter is so large for those physical quantities that are apparently nothing to do with fine-tuning.

\section{Explanations}

To further investigate the possible intrinsic connections between the problems of the sensitivity parameter and those consequences\cite{Dent,Calmet,Langacker,Dine} we need to explain if the fine structure constant is time varying, it is necessary to revisit the possible mechanisms involved in comparing a dimensionful mass parameter with a dimensionless coupling constant.

Suppose we have a general quantum field theory with a dimensionless coupling constant \(g\) and several dimensionful mass parameters \(m_i\), without considering any specific interactions, the lowest order renormalization group equations can be written as\cite{Pich,MV1,MV2,MV3,Luo}:

\begin{equation}
\frac{dm_i}{dt}=\gamma_{ij}(g)m_j+\cdots
\label{eqx10}
\end{equation}

\begin{equation}
\frac{dg}{dt}=\beta(g)+\cdots
\label{eqx11}
\end{equation}
where \(t=\ln \Lambda/\Lambda_0\).  \(\beta\) and 
\(\gamma_{ij}\) are dimensionless functions of the coupling constant. Because \(g\) is dimensionless so the function \(\beta(g)\) here can not have the first order term.  As an example, in QCD they are\cite{MV1,MV2,MV3,Luo}:
\begin{equation}
\beta(g)=-b\frac{g^3}{16\pi^2}+...
\label{eqx12}
\end{equation}

\begin{equation}
\gamma_{ij}=\gamma^0_{ij}\frac{g^2}{16\pi^2}+...~~(\gamma^0_{ij}=0~~\mbox{if}~~i\ne j)
\label{eqx13}
\end{equation}
The solutions of Eq.~\ref{eqx10} and Eq.~\ref{eqx11} can be written in the following matrix formation:
\begin{equation}
\mathbf{m}=\mathbf{m}_0e^{\int_{t_0}^{t}\mathbf{\Gamma}dt}+\cdots
\label{eqx14}
\end{equation}
where  \(\mathbf{m}=(m_i)\), \(\mathbf{\Gamma}=(\gamma_{ij})\) are matrices. The matrix
\(\mathbf{m}_0\) is the initial value of the matrix \(\mathbf{m}\), obviously it does not explicitly depend on the running of the energy scale. From the renormalization point of view, generally a physical quantity(for example, \(\mathbf{m}\) in Eq.~\ref{eqx14}) can be separated into two factors: the factor that explicitly depends on   the running of the energy scale(for example, \(\exp(\int_{t_0}^{t}\mathbf{\Gamma}dt)\) in Eq.~\ref{eqx14}), and the factor that is renormalization invariant(for example, \(\mathbf{m}_0\) in Eq.~\ref{eqx14})\cite{Dent}. This separation is very important, because in renormalization the factor that depends on the running of the energy scale mainly decides whether the physical quantity is fine-tuned or not. 

Here we take QCD as an example. The reason why we choose QCD is simply
because fermion masses in QCD are protected by the gauge symmetry, they are not fine-tuned.
Therefore we can easily judge whether a fine-tuning measure is correct or not. 
For simplicity assume Eq.~\ref{eqx10} and Eq.~\ref{eqx11} only have one mass parameter \(m\),
as a result of the gauge symmetry, the one loop correction to the mass parameter will diverge logarithmically rather than quadratically:
\begin{equation}
\delta m \approx g^2\ln{\frac{\Lambda}{\Lambda_0}}
\end{equation}
Obviously the logarithm function removes the fine-tuning possibility. Solving  Eq.~\ref{eqx10} and Eq.~\ref{eqx11}, the solution will be\cite{Pich}:
\begin{equation}
m(t)=m(t_0)\Biglb(\frac{g(t)}{g(t_0)}\Biglb)^{-\gamma^0/b}
\label{eqx15}
\end{equation}
The sensitivity for the renormalization dependent factor \((g(t)/g(t_0))^{-\gamma^0/b}\) will be:
\begin{equation}
c\approx\gamma^0/b
\label{eqx16}
\end{equation}
Because the anomalous dimensions are usually small\cite{pdg}, thus the result of Eq.~\ref{eqx16} usually is much smaller than \(10\). For this part, the sensitivity parameter is consitent with the severity of fine-tuning.

As for the renormalization invariant factor \(m(t_0)\), here it is just an initial value and does not contribute to the fine-tuning. Therefore in renormalization mainly the factor that explicitly depends on the energy scale controls whether a physical quantity is fine-tuned or not. If the sensitivity of the  renormalization invariant factor \(m(t_0)\) is also neglectable, then the severity of fine-tuning will be consistent with the largeness of the sensitivity parameter. If everything ends up here, then we won't have any problem of the sensitivity parameter.

As X.Calmet and H.Fritzsch\cite{Calmet} and Langacker et al\cite{Langacker} pointed out, due to the existence of the asymptotic freedom, the QCD scale parameter \(\Lambda_{QCD}\) will be highly sensitive to the strong coupling constant:

\begin{equation}
\frac{\delta\Lambda_{QCD}}{\Lambda_{QCD}}=38\frac{\delta \alpha_s}{\alpha_s}
\end{equation}

Besides, due to the coupling constant unification, any energy scale parameter will also be highly sensitive to the variation of a gauge coupling constant\cite{Calmet,Langacker}:

\begin{equation}
\frac{\delta\Lambda}{\Lambda}=(38\pm6)\frac{\delta \alpha}{\alpha}
\label{eq18}
\end{equation}

Becuase for any dimensionless couplings, the lowest order term of its \(\beta\) function is always the third order term\cite{MV1,MV2,MV3,Luo}, therefore, the mathematical relation between a dimensionless coupling and the energy scale parameter \(\Lambda\) must be an exponential function, which means the energy scale parameter \(\Lambda\) is always highly sensitive with respect to  minute variations of the dimensionless coupling constant. 
Therefore in Eq.~\ref{eq18} if the fine structure constant \(\alpha\) is replaces by another  dimensionless coupling, we will still have the similar result. Apparently,
the mass dimension difference between the dimensionless coupling and the dimensionful energy scale parameter \(\Lambda\) is the origin of these large sensitivities.  This phenomenon can be treated as  one of the intrinsic properties of the energy scale parameter and should not be understood as a theoretical problem. The large sensitivity given by Eq.~\ref{eq2} obviously does not mean fine-tuning.

The renormalization invariant factor of a physical mass usually is its initial value at a given energy scale.  Therefore principally, if there exists a mechanism that ends up with a linear relation between a specific mass and \(\Lambda\), and if that specific mass is used to define the renormalization invariant factor, then we will be able to find a relation similar to
\begin{equation}
m(t_0)\sim\Lambda
\label{eq19x}
\end{equation} 
if this is true, then the  sensitivity parameter of \(m(t)\) to \(g\) will be: 
\begin{equation}
c=\frac{\gamma^0}{b}+\frac{g(t_0)}{\Lambda}\frac{\partial\Lambda}{\partial g(t_0)}\simeq\frac{g(t_0)}{\Lambda}\frac{\partial\Lambda}{\partial g(t_0)}\simeq38
\label{eq20x}
\end{equation}
The sensitivity given by Eq.~\ref{eq20x} is completely different from the sensitivity given by Eq.~\ref{eqx16}. 
The large sensitivity of \(\Lambda\) to \(g\) from Eq.~\ref{eq18} greatly changes the sensitivity of \(m(t)\) to \(g\).
Based on the sensitivity criterion, the mass \(m\)  is highly fine-tuned, even though it is protected by the gauge symmetry and is nothing to do with fine tuning.  This example shows why the sensitivity parameter failed in the cases discussed in the introduction. 
 
In particle physics, Many mechanisms can give a relation similar to Eq.~\ref{eq19x}. For example, for theories that are asymptotically free, the running of the coupling constant \(g\) can produce a mass \(m(g,\Lambda_0)\) via dynamical symmetry breaking\cite{Gross}:

\begin{equation}
m(g,\Lambda_0)=\Lambda=\Lambda_0 e^{-\int dg/\beta{g}}
\label{eqx17}
\end{equation}

If the initial value of a physical mass (the renormalization invariant factor) is given by Eq.~\ref{eqx17}, then apparently the physical mass will also be highly sensitive to \(g\), even if the mass quantity itself has been stabilized by the gauge symmetry. 
The mechanism of the dynamical symmetry breaking is widely used, therefore many mass parameters will be affected. For example, in supersymmetric standard models, dynamical symmetry breaking is used to specify the initial values of the soft masses\cite{Witten}, so soft masses will be highly sensitive to \(g\). Similar examples have already been 
studied by P. Langacker et al.\cite{Langacker} and B. Campbell et al.\cite{Campbell} for the time-varying coupling constant problems.

Besides the dynamical symmetry breaking, other mechanisms such as gaugino condensation, chiral symmetry breaking
etc also have the same effect. For example, chiral symmetry breaking\cite{Gellmann} gives the following relation:
\begin{equation}
\langle0|q\bar{q}|0\rangle^{1/3}\sim \Lambda
\label{eqx177}
\end{equation}
If the initial value of a nucleon mass is defined by Eq.~\ref{eqx177}, then the nucleon mass will be highly sensitive to the variation of \(g\). This scenario is similar to the time-varying coupling constant problems discussed by T. Dent et al.\cite{Dent} and T. Chiba et al\cite{Chiba}.

Technically, the problem of the sensitivity parameter is the counterpart of the problem that related with the time-varying coupling constants.  The time variations of the fine structure constant \(\dot{\alpha}\) corresponds to the variations of the initial value \(\delta x\) here. Similar to our analysis, the large time variations of the physical quantities are caused by the large time variations of the energy scale parameter \(\dot{\Lambda}\), and the large time variations of the energy scale parameter  \(\dot{\Lambda}\) are caused by the small time variations of the fine structure constant \(\dot{\alpha}\). The last step is the consequence of the  mass dimension difference between the fine structure constant \(\alpha\) and the energy scale parameter \(\Lambda\). The large relative changes of nucleon mass and many other dimensionful quantities in the time-varying coupling constant problem are obviously due to the native scale difference between the energy scale parameter and the dimensionless coupling constants. 

We can easily verify that either the problems of the sensitivity parameter, or the problems related with the time-varying coupling constant disappear when two physical quantities compared have the identical mass dimension. Replace the mass matrix \(\mathbf{m}\) with two dimensionful mass parameters \(m_i\) and \(m_j\), the solution of Eq.~\ref{eqx10} now becomes\cite{Pich}:

\begin{equation}
m_i(t)=\sum_{j,k}U_{ij}e^{\int dt(-\mathbf{\Gamma}^0/b)}U^{-1}_{jk}m_k(t_0)
\label{eqx20}
\end{equation}
where \(\mathbf{\Gamma}^0\) is the matrix formation of \(\gamma^0_{ij}\), \(U_{ij}\) is the element of the matrix that diagonalizes \(\mathbf{\Gamma}^0\).

Because in Eq.~\ref{eqx20},  the exponent \(\mathbf{\Gamma^0}/b\) does not explicitly contain any  mass parameter, thus the value of \(\partial{m_i(t)}/\partial{m_j(t_0)}\) will not be affected 
by the factor \(\partial{\Lambda}/\partial{g}\). So it is quite straightforward to conclude that 
the sensitivity of a dimensionful mass parameter \(m_i(t)\) to another dimensionful mass parameter \(m_j(t_0)\)  won't have the problem. Because of the mathematical similarities, we can also conclude that the problems caused by the time-varying coupling constant also only exist in the dimensionful parameters. The dimensionless parameters won't have such problems.

\section{Discussions}

There are several important consequences for this mechanism. First, it affects the validity of the sensitivity criterion. Certainly this doesn't mean the sensitivity criterion will fail under any circumstances. As we have analyzed, the sensitivity criterion is unreliable only when a dimensionful quantity and a dimensionless coupling constant are compared. And furthermore, these two parameters need to be at two different energy scales, thus need to be mathematically linked by renormalization. Besides, there should be a mechanism which mathematically associate the renormalization invariant factor of the dimensionful mass quantity to the energy scale parameter \(\Lambda\).

If the above conditions are not met, then the sensitivity parameter will not fail. For example, for the fine-tuning problems exist in the mass mixing\cite{Casas:2004gh,AD}, where all parameters involved  are at the same energy scale and a renormalization evolution is not required to mathematically link these parameters. Therefore the sensitivity parameter can be used as an accurate measure for the severity of fine-tuning. Estimate the severity of fine-tuning by comparing two parameters with different mass dimensions at different energy scales is the  situation that we need to pay special attention to.

To solve these problems, many researchers have proposed many explanations and alternative prescriptions\cite{GWA,AC2,AC3,CS,BR,BS,GRS,RS,Chan, Chankowski, Chankowski2,Kane, Gil}. Those alternative fine-tuning 
measures received severe criticism\cite{Abe:2001npb,Feng}. Obviously, a good fine-tuning
measure can not be properly defined without the knowledge of which mechanism causes the failure of the sensitivity measure. Because otherwise the results based on these alternative prescriptions could also be misleading. 

Based on the analysis in the previous section, we now finally know which
mechanisms can fail the sensitivity measure. But due to the complexity of the real models, 
the effects of these mechanisms to the sensitivity measure are highly model dependent. It is quite
impossible to remove these effects and then invent a simple yet universal prescription that can be easily applied to any models. The advantages of the sensitivity criterion are obvious, though occasionally it has problems. If we want to find a way to measure the fine-tuning correctly without losing the beauty of simplicity and straightforwardness, it is better to keep the sensitivity criterion but avoid using it when two parameters with different mass dimensions at different energy scales are compared, because this is the simplest and the most efficient solution for these problems.

Second, because the mechanisms such as the dynamical symmetry breaking, the gaugino condensation etc play important roles in modern particle physics theories. While experimentally, we still have difficulties to directly verify the existence of these mechanisms. The time variation problem discussed here connects those mechanisms with the astrophysical observation, therefore it can be used as a probe to test new particle physics theories  
without resorting to a powerful accelerator. If in the future we can find an astrophysical way to measure the time variations of the proton mass or any other dimensionful physical quantities, for example, using Zeeman effect\cite{Zeeman} to measure the mass-to-charge ratio in astrophysical observation, we may be able to learn the fermion masses at the early cosmological epoch. By comparing the variations with our theoretical predictions, we can indirectly learn the details of these mechanisms, for example, the gaugino condensation. 

Finally, as we mentioned in the first part, the fine-tuning cut-off sensitivity \(c=10\) is an arbitrary value manually chosen by Barbieri and Giudice. We don't have any experimental evidence to support this value. Physical quantities with \(c\ge10\) could be fine-tuned but could also be nothing to do with fine-tuning. If in some way we can learn the time variations of various physical quantities by the astronomical observation, then we can further calculate the real sensitivities of these physical quantities in our universe.  With these data we will be able to choose a more reasonable fine-tuning cut-off for the sensitivity parameter,  not just the arbitrary value \(c_{\text{max}}=10\) chosen\cite{BG}.

\section{Conclusions}
The problems of the sensitivity parameter have been discovered for more than ten years. Many analyses, explanations and alternative prescriptions have been proposed to solve these problems. But none of them has been accepted widely. The reason why the sensitivity parameter fails under specific circumstances still remains unclear.

The recent discovered time-varying fine structure constant and its related problems give us a clue to deal with the problems of the sensitivity parameter. We have investigated the examples that the sensitivity parameter fails, compared them with the problems associated with the time-varying coupling constants, demonstrated that the reason why the sensitivity parameter fails to represent the true level of fine-tuning is because the renormalization invariant factor of a dimensionful parameter is linked to \(\Lambda\) via various mechanisms. Thus the large sensitivity of \(\Lambda\) to the dimensionless coupling will greatly influence the sensitivity calculation, severity of fine-tuning will be over-estimated. Technically, the problems associated with the sensitivity parameter are identical to the problems associated  with the time-varying coupling constants.

Finally, we want to point out that this effect only exists when parameters with different mass dimensions at different energy scales are compared. Based on these analyses, we argue that the best way to avoid these problems is always compare parameters with same mass dimensions if they are at the different energy scales.

\end{document}